\documentclass[aps,pra,reprint,groupedaddress,twocolumn]{revtex4-2}
\usepackage{graphicx}
\usepackage{xcolor}
\usepackage{amsmath}
\usepackage{mathtools}
\usepackage{amssymb}
\DeclareMathOperator*{\argmax}{arg\,max}
\newcommand{\bs}[1]{\boldsymbol{#1}}
\newcommand{\mc}[1]{\mathcal{#1}}
\newcommand{\defeq}{\vcentcolon=}

\begin{document}

\title{Quantum-enhanced pattern recognition }
\author{Giuseppe Ortolano$^{1,2}$}
\author{Carmine Napoli$^1$}
\author{Cillian Harney$^3$}
\author{Stefano Pirandola$^3$}
\author{Giuseppe Leonetti$^{1,2}$}
\author{Pauline Boucher$^1$}
\author{Elena Losero$^1$}
\author{Marco Genovese$^1$}
\author{Ivano Ruo-Berchera$^1$}
\affiliation{$^1$Quantum metrology and nano technologies division, INRiM, Strada delle Cacce 91, 10153 Torino, Italy}
\affiliation{$^2$DISAT, Politecnico di Torino, Corso Duca degli Abruzzi 24, 10129 Torino, Italy}
\affiliation{$^3$Department of Computer Science, University of York, York YO10 5GH, United Kingdom}

\begin{abstract}
The challenge of pattern recognition is to invoke a strategy that can accurately extract features of a dataset and classify its samples. In realistic scenarios this dataset may be a physical system from which we want to retrieve information, such as in the readout of optical classical memories.
The theoretical and experimental development of quantum reading has demonstrated that the readout of optical memories can be dramatically enhanced through the use of quantum resources (namely entangled input-states) over that of the best classical strategies. However, the practicality
of this quantum advantage hinges upon the scalability of quantum reading, and up to now its experimental demonstration has been limited to individual cells. In this work, we demonstrate for the first time quantum advantage in the multi-cell problem of pattern recognition. Through experimental realizations of digits from the MNIST handwritten digit dataset, and the application of advanced classical post-processing, we report the use of entangled probe states and photon-counting to achieve quantum advantage in classification error over that achieved with classical resources, confirming that the advantage gained through quantum sensors can be sustained throughout pattern recognition and complex post-processing. This motivates future developments of quantum-enhanced pattern recognition of bosonic-loss within complex domains. 
\end{abstract}

\maketitle
 
Significant progress has been made in recent years in quantum sensing \cite{Degen_2017,Pirandola_2018,Petrini_2020}, a broad spanning field of research which seeks to realise quantum protocols for estimation \cite{Helstrom_1976,Braunstein_1994,Braunstein_1996,Giovannetti_2004,Giovannetti_2011} and discrimination \cite{Helstrom_1969,Chefles_1998,Chefles_2000,Bergou_04} that achieve better performance than any classical protocol, i.e.~gain a quantum-advantage. Efforts to design and engineer such protocols have been widely successful, with applications in data-readout \cite{Pirandola_2011, Pirandola_2011a, Nair_2011, Invernizzi_2011, Dallarno_2012, Spedalieri_2012,Tej_2013, Ortolano_2021, Ortolano_2021b, Ortolano_2022}, target detection \cite{Sacchi_2005, Lloyd_2008,  Tan_2008, Lopaeva_2013, Zhang_2013,Zhang_2015, Barzanjeh_2015, Sanz_2017,  Zhuang_2017a,Zhuang_2017b,Barzanjeh_2020, Casariego_2022, Zhuang_2022b, GonzalezRaya_2022}, cryptography \cite{Spedalieri_2015,Pereira_2021a}, fundamental physics research \cite{Brady_DM_2022,Marchese_2021, Berchera_2013} and imaging  \cite{Genovese_2016, Losero_2018, Spedalieri_2020}.

An application of particular interest is that of quantum reading, which encompasses the challenge of classical information retrieval from an optical memory. A memory cell can be considered as a reflective (or transmissive) target with two distinct possible reflectivities (transmittivities) encoding a classical bit. This represents a pair of possible bosonic-lossy channels that act on incident photonic probe states. Accurate readout of a memory cell requires the correct discrimination of output states from the channel (i.e.,~quantum channel discrimination). Having conceptualised this model, Pirandola \cite{Pirandola_2011} proved that one can significantly enhance the accuracy of information retrieval through the use of entangled light sources. This seminal work led to a series of advancements and most recently, a successful experimental demonstration \cite{Ortolano_2021}.

The success of quantum reading prompted the development of further schemes in more complex domains, such as the simultaneous readout of many memory cells. This breeds the task of quantum multi-channel discrimination \cite{Zhuang_2020,Zhuang_2020a}, in which one must optimise multi-mode probe states and measurements in order to classify a collection of bosonic-lossy channels (or \textit{channel patterns}). This development brings new layers of complication; expanding the space of possible channel patterns to be discriminated, and revealing a more general spectrum of multi-modal discrimination protocols. This invites new applications in which quantum reading may serve as a primitive. 

The use of quantum sensors within the task of pattern recognition provides such an opportunity, where one is tasked with classifying a set of patterns (or images). The pattern distribution is often non-uniform, and classifications follow highly non-linear relationships, e.g.~the classification of hand-written digits. Powerful classical post-processing techniques are well known for this challenge, however the quality of the readout of data-sets are often taken for granted. Given a system that can be modelled as an ensemble of bosonic-lossy multi-channels, it has been theoretically shown that quantum reading can substantially enhance classification power over optimal classical strategies \cite{Banchi_2020,Pereira_2020,Harney_2021a,Harney_2021b,Harney2021c}.

In this work, we report the experimental demonstration of quantum-enhanced pattern recognition using signal-idler entangled two-mode squeezed vacuum  (TMSV)  states and photon-counting measurements to outperform any classical strategy. Using the MNIST handwritten digit dataset \cite{MNIST}, image samples are experimentally realised and serve as a test-bed for the discrimination protocols. Investigating both multi-shot and single-shot regimes, we show that the use of entangled probe states can generate quantum advantage in the classification error as a consequence of the gain offered by quantum resources. By considering multiple supervised-learning approaches to classification, we emphasise that this advantage is an inherent property of quantum sensing that can survive throughout complex post-processing. Our work paves the way for a suite of near-term applications of quantum-enhanced pattern recognition and imaging.  

\section{Results}

\begin{figure}
	\vspace{0.10cm}
	\includegraphics[width=1\columnwidth]{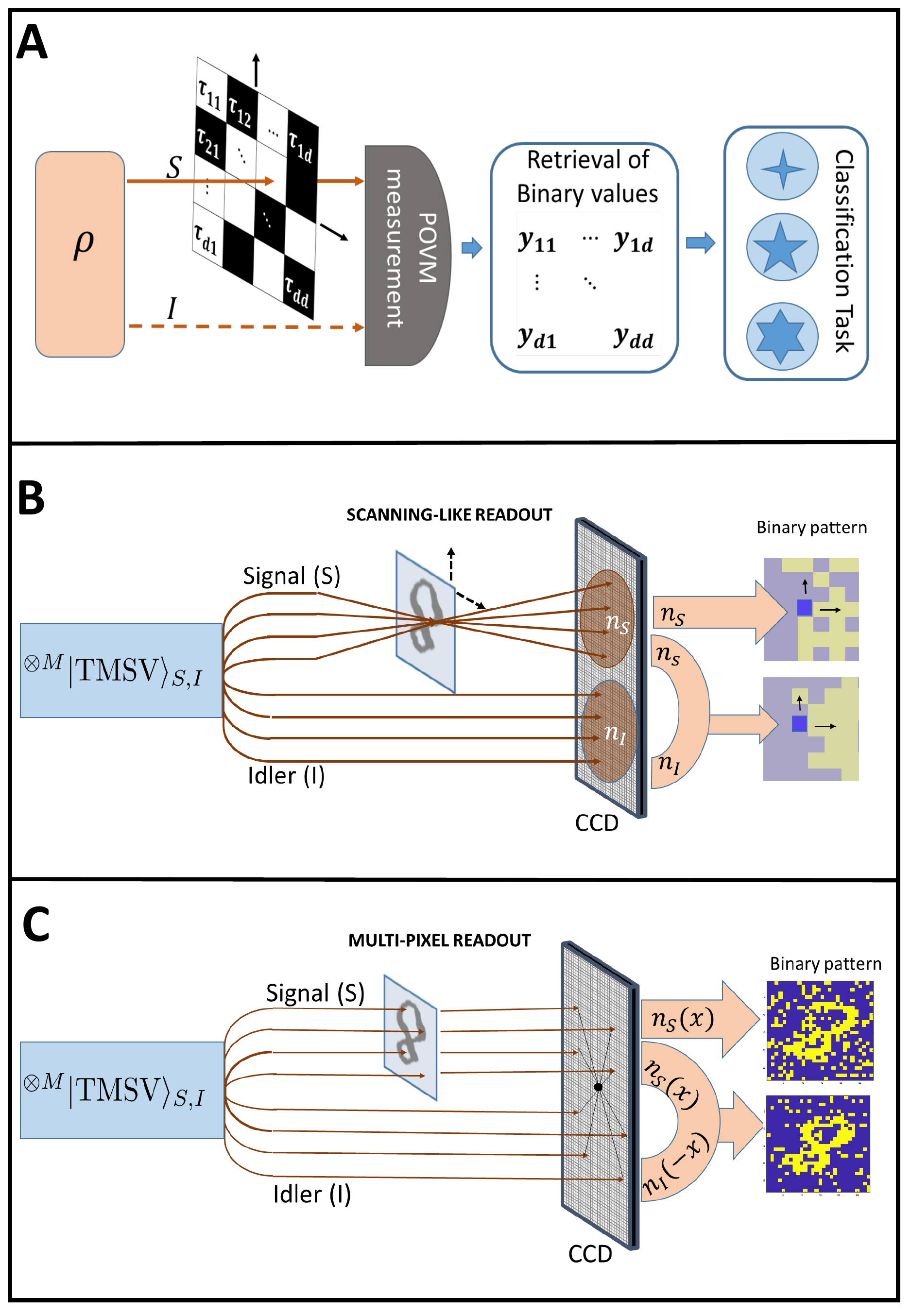}
	\caption{\emph{Scheme of Pattern Recognition}. Panel \textbf{A} shows the general configuration of a pattern recognition problem. An image composed of binary pixels, having one of two possible values of transmittance $\tau_{i,j}=\{\tau_{0},\tau_{1}\}$, is probed by a bipartite state $\rho$. The signal (S) system interacts with the image. The idler (I) system is directly sent to the measurement apparatus where a joint measurement is performed with the signal. The result of the measurement are processed classically to classify the patterns. Panel \textbf{B} shows a possible experimental configuration, in which the pattern is "scanned" with sequential single-pixel PC measurements, where $n_{S}$ and $n_{I}$ are the signal and idler numbers of photons detected. Panel \textbf{C} shows a different configuration in which the spatial multi-mode structure of the source is used, so that a parallel readout of all the pixels of the image is done in a single shot. The pairwise quantum correlation of the different modes, represented schematically by the brown arrows hitting symmetric points of the CCD,a are used to enhance the classification.}\label{Scheme}
\end{figure}

Let us consider a pattern classification problem where spatial patterns are imprinted on a $d\times d$ array of binary cells, each encoding one bit of information in two values of a physical parameter, for example the transmittances $\tau_0$ and $\tau_1$, where we pose $\tau_0\leq\tau_1$ (see Fig. \ref{Scheme}\textbf{A}). The cells size and dimension of the array $d^2$, can be a native characteristic of the physical support but also they can be practically determined by resolution properties of the sensing apparatus, for example the pixel structure of a camera or the resolution of an optical system, so that $\tau_0$ and $\tau_1$ are intended as the values corresponding to each readout pixel. The sensing strategy has to be designed to minimize the average classification error probability of the patterns, $P$.

The general sensing procedure can be described as in Fig. \ref{Scheme}\textbf{A}: an input state, represented by the density operator $\rho$, interacts with the object encoding the pattern, and a POVM measurement is performed on the output state, followed by classical post-processing. The state $\rho$ is generally bipartite in order to include ancillary assisted schemes. The two systems are indicated as signal (S) and idler (I), according to quantum optics terminology. Since we are dealing with bosonic systems, the problem is non-trivial only if we impose some constraints on the resources, the most natural one being to fix the mean number of signal photons, $\mu$, probing the cell. The pixels are assumed to be probed by independent spatial modes, and in this work we restrict our analysis to local measurements, i.e. independent readout is performed for each pixel at the receiver.

Following the local measurement, a binary value is assigned to each pixel with a certain average probability of error $p$. The binary image is then classified. The classification error $P$ is related to the bit-flip probability $p$ by a, generally, non-linear map $\mathcal{M}_C$:
\begin{equation}
P=\mathcal{M}_C(p).
\end{equation}
The map $\mathcal{M}_C$ depends on the problem at hand, but it is reasonable to assume it monotonic for any effective classification algorithm. 
Using the monotonicity of $\mathcal{M}_C$, the minimization of $P$ is obtained by minimizing $p$ over all possible input states and POVM measurements. Details on the single-pixel problem are provided in Sec.~\ref{sec:SinglePix}.

The theoretical lower bound on the error probability achievable by classical states, $p^{\mathtt{LB}}_{\mathtt{cla}}$ is reported in Eq.~(\ref{eq:2}). It can be saturated, without idler assistance, by a single mode coherent state $|\alpha\rangle$, and an optimal measurement consisting in the projection on two pseudo-cat states as shown in Sec.~\ref{OptMeas}.

Such a measurement appears to be experimentally unfeasible and no other receiver that saturates $p^{\mathtt{LB}}_{\mathtt{cla}}$ has been proposed so far.
Nevertheless, a photon counting (PC) receiver, composed of PC measurement followed by a Bayesian decision, can be proven to perform close to the optimal one for practical purposes \cite{Ortolano_2021}. As the PC is a phase-insensitive measurement, it follows that any state with Poisson photon number distribution can reach the PC limit, $p^{\mathtt{PC}}_{\mathtt{cla}}$, as reported in Eq.~(\ref{eq:4}). 
For this reason, we use a PC receiver to experimentally perform the pixel readout with both classical and quantum states. 

The quantum readout was performed using a collection of $M$ replicas of TMSV states, $\rho_{\mathtt{qua}}=|\text{TMSV}\rangle_{S,I} \langle \text{TMSV} |^{\otimes M} $, with

\begin{equation}
	\vert \text{TMSV}\rangle_{S,I}\propto\sum_{n} \sqrt{\mathrm{P}_{\mu_0}(n)}\vert n\rangle_{S}\vert n\rangle_{I}.
\end{equation}\label{TMSV}
$\mathrm{P}_{\mu_0}(n)=\mu_0^n/(\mu_0 + 1)^{n+1}$ is the thermal distribution with $\mu_0=\mu/M$ mean photons. In the quantum case, idler assistance is crucial and it is actually convenient to spread the signal energy across a large number of modes, i.e. to work with $M\gg1$ and $\mu_0\ll1$ \cite{Ortolano_2021}. The corresponding error probability $p^{\mathtt{PC}}_{\mathtt{qua}}$, discussed in Sec. \ref{sec:SinglePix}, can be evaluated numerically.

Among the various experimental imperfections, the most relevant is photon loss due to non-ideal optical component and detector inefficiency. The effect of the losses on classical readout performances is equivalent to a rescaling of the energy, namely the input mean photon number is substituted by the detected photons, $ \mu\longrightarrow\eta\mu $ in the classical bounds of Eqs.~(\ref{eq:2}) and (\ref{eq:4}), where $\eta$ is the overall photon detection probability. However, the quantum strategy is additionally affected by deterioration of the bipartite non-classical correlation due to losses, which describes the imperfect coupling of correlated modes into the detectors. This effect is discussed in Sec. \ref{sec:multipix}.

\subsection{Experimental realization}

The principle of the experimental scheme is presented in Fig. \ref{Scheme} \textbf{B}-\textbf{C}. The $M$-product TMSV quantum state $\rho_{\mathtt{qua}}$ is produced by traveling wave Spontaneous Parametric Down Conversion (SPDC) of type II in a non-linear crystal pumped by a continuous laser. Due to the conservation of the transverse momentum of the pump $\textbf{q}=0$, it follows $\textbf{q}_S=-\textbf{q}_I$, i.e. signal and idler photons are emitted along correlated directions (spatial modes), which are mapped into symmetric pixels of a CCD camera. In our experiment the exposure time of image acquisition is much longer than the coherence time of the process, so many temporal modes, on the order of $10^{11}$, contribute to the total number of photons detected in each pixel (few thousands). Clearly, the conditions $M\gg1$ and $\mu_0\ll 1$ which maximize the quantum advantage are fulfilled and the single pixel photon statistics is practically indistinguishable from the Poisson one. For this reason, the best classical performance with PC detection can be evaluated by considering only the signal beam. 
Details on the experimental setup are reported in Sec. \ref{sec:ExpSet}.
 
In general, traveling wave SPDC is spatially broadband. There are many (several thousands) pairwise correlated spatial modes available for spatially resolved multi-pixel quantum readout \cite{Brida_2010, Samantaray_2017}.
However, a trade-off should be set between the spatial resolution, in terms of pixels available in one shot measurement, and the reduction of the pixel error probability due to quantum correlation. This point is discussed later on in the main text, and further details can be found in Sec. \ref{Materials&Methods}.

To address this trade-off, we have realized the experiment in two different configurations, one named ``scanning-like readout'' (Fig.~\ref{Scheme}\textbf{B}) that favors the best quantum enhancement, and the second one named ``multi-pixel readout'' (Fig.~\ref{Scheme}\textbf{C}) that allows spatially resolved pattern acquisitions with in a single-shot measurement. 

\subsection{Scanning-Like Readout}

\begin{figure}
	\vspace{0.10cm}
	\includegraphics[width=\columnwidth]{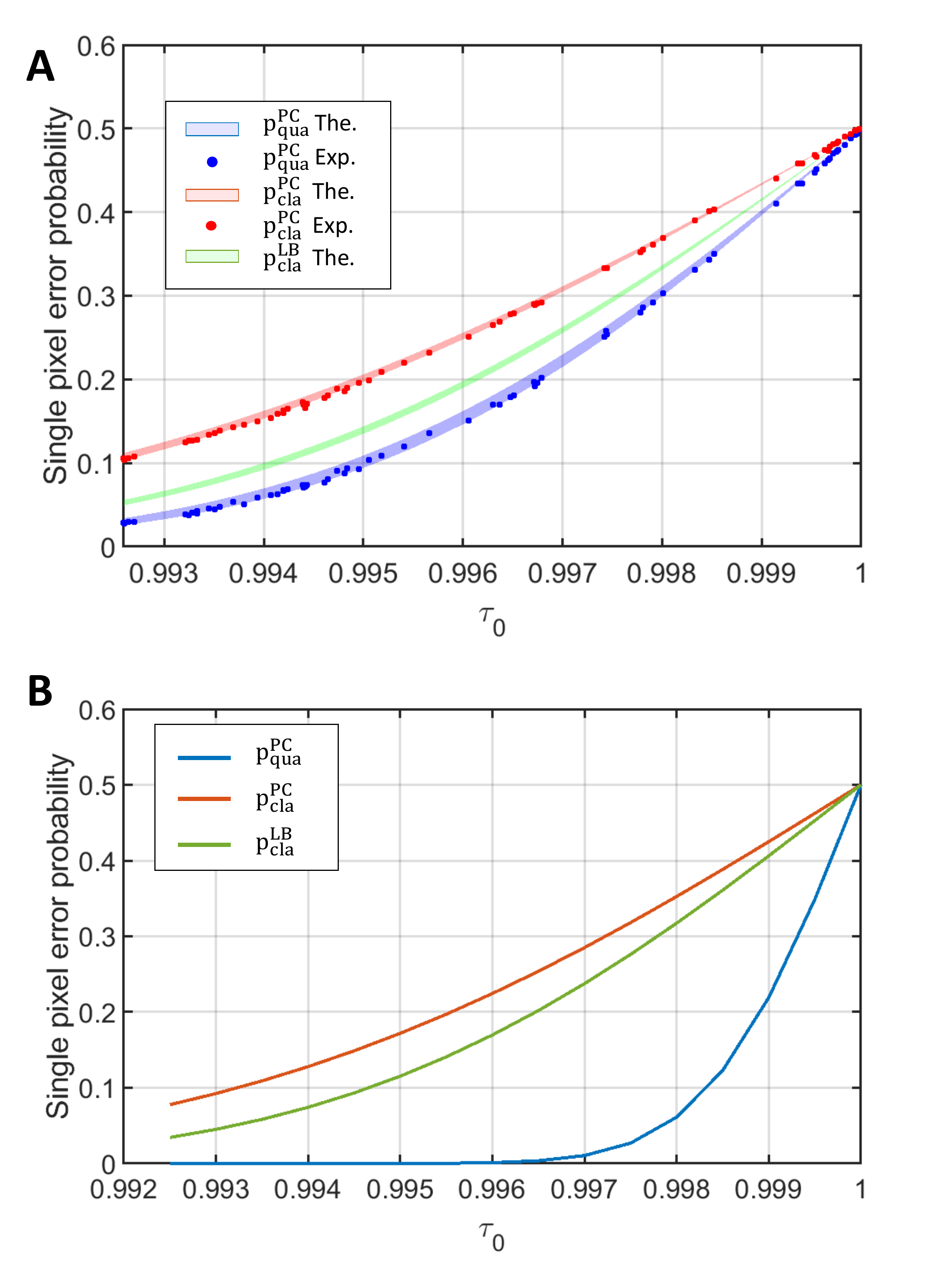}
	\caption{\textit{Single-pixel readout error probabilities.} Panel \textbf{A} reports the error probability in the readout of a bit  encoded in the transmittance value (either $\tau_0$ or $\tau_1$) of a single-pixel. The lower transmittances, $\tau_0$, is varied  while the other one fixed to $\tau_1=1$. The estimated mean number of photons per pixel is $\mu=(1.45 \pm 0.05) \cdot 10^5$, the efficiencies are $\eta_S=0.79 \pm 0.01$ and $\eta_I=0.791 \pm 0.004$, and the electronic noise is $\nu\approx 6 \cdot 10^3$. The probability of error for the quantum strategy is reported in blue, the classical strategy with photon counting in red and the best classical strategy in green. The confidence interval of the theoretical curves is obtained by considering one standard deviation on the reported parameters. Panel \textbf{B} reports theoretical prediction on the error probabilities in the almost ideal case of $\eta_S=\eta_I=0.97$, while all the other parameters are the same of panel \textbf{A}.}\label{PxErr_X2}
\end{figure}

In this first experimental configuration, we mimic a readout in which the physical binary pattern encoded by the transmittance $\tau_0$ and $\tau_1$ is scanned point-by-point, as shown in Fig.~\ref{Scheme}\textbf{B}. 
First, we repeated a large number of independent measurements of two well-characterized transmittances, $\tau_0$ and $\tau_1$, assigning a bit value $ (y=0,1)$ for each measurement according to the Bayes' rule in Eq. (\ref{Bayes}) based on the number of photon-counts $n_S$ and $n_I$ (see Sec. \ref{Materials&Methods}A). The photon-counts are integrated over two large regions of the CCD array, defining the size of the readout pixel, collecting a large number of signal and idler spatial modes respectively, obtaining the best possible correlation between $n_S$ and $n_I$, as discussed in Sec. \ref{Materials&Methods}D. This produces two sets of data $\textit{A}_0=\{y^{(0)}_k\}_{k=1,..,K}$ and $\textit{A}_1=\{y^{(1)}_l\}_{l=1,..,L}$ for the nominal transmittance $\tau_0$ and $\tau_1$ respectively, with $L+K\gg d^2$. The procedure is performed both for classical (only using signal beam) and quantum reading. From those sets we can evaluate the average experimental readout pixel error probabilities, $p^{\mathtt{PC}}_{\mathtt{cla}}$ and $p^{\mathtt{PC}}_{\mathtt{qua}}$ shown in Fig.~\ref{PxErr_X2}\textbf{A}. One of the transmittance is fixed to $\tau_1=1$ while the other one, $\tau_0$, is varied. All other parameters are kept fixed. The experimental points for both the quantum (blue) and classical (red) readout are compared with the classical lower bound $p^{\mathtt{LB}}_{\mathtt{qua}}$(green) defined by Eq.~(\ref{eq:2}). As expected, in general the error probability increases as the two transmittances become closer and closer. However, through the range of parameters showed, the strategy employing quantum states sensibly outperforms the classical one paired with the suboptimal PC measurement and is also able to surpass the absolute classical lower bound. For the sake of completeness, in Fig. \ref{PxErr_X2}\textbf{B} we report the same pixel error probabilities in an almost ideal scenario where the detection efficiency is fixed to $\eta=0.97$ (the efficiency reached in the actual setup is $\eta\approx 0.79$). 

Note that, the data picked from the set $\textit{A}_0$ and $\textit{A}_1$ can be virtually rearranged to reproduce the outcomes of a sequential spatial scanning of any given $d\times d$ binary pattern. However, for an accurate evaluation of the pattern recognition task, the used approach is highly preferable with respect to the real scanning of spatial samples. In fact, the last one would require the physical realization and scanning of a large number of spatial samples, which is hardly feasible and actually not necessary.

\begin{figure}
	\vspace{0.10cm}
	\includegraphics[width=1.0\columnwidth]{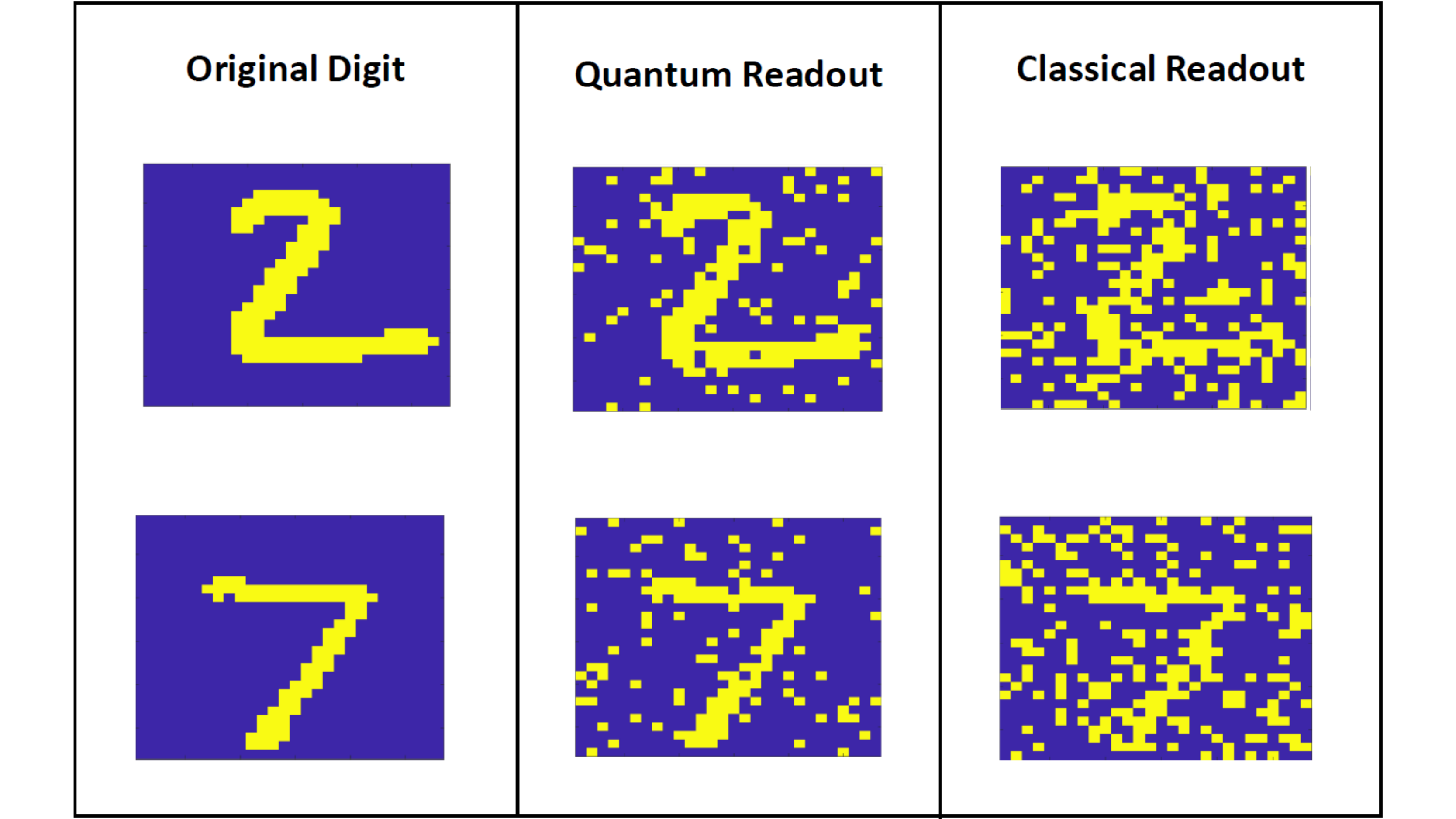}
	\caption{\textit{Readout of handwritten digits with different strategies.} Two examples of binary pattern of dimension 28x28 with quantum and classical PC readout noise .}\label{Patterns}
\end{figure}

To evaluate the pattern recognition performance we used the MNIST handwritten digit dataset, containing 60000 training samples of handwritten numerical digits and 10000 test samples. We assemble the entire set of test patterns by picking experimental data from the set $\textit{A}_0$ and $\textit{A}_1$. Fig.~\ref{Patterns} shows some examples of digit contained in the dataset, and a comparison of the effect of the noise in either a classical and a quantum binary readout.

\begin{figure*}
	\vspace{0.10cm}
	\includegraphics[width=1\linewidth]{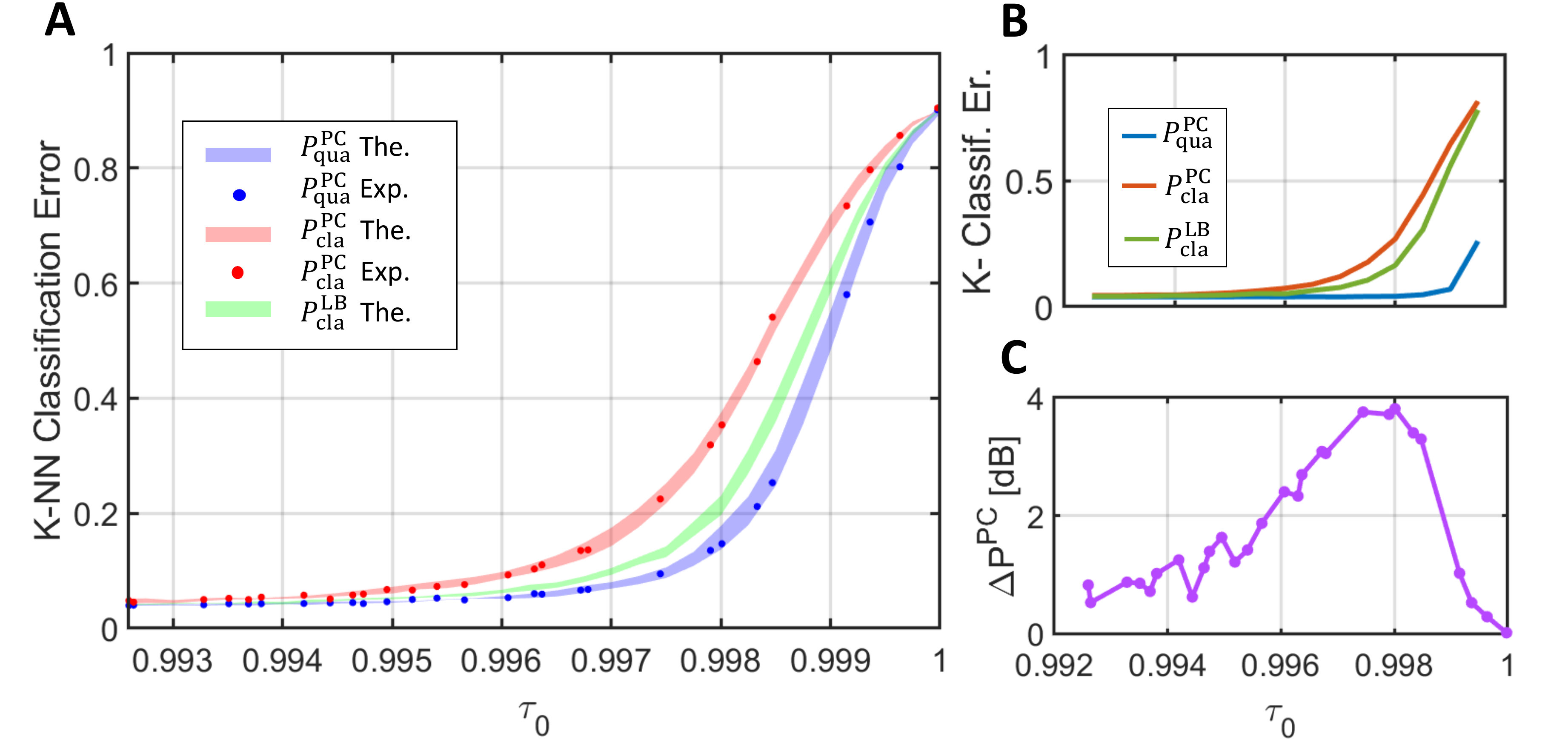}
	\caption{\textit{Classification error with the $k$-NN algorithm}. Panel \textbf{A} shows the average classification error $P$ in the recognition of handwritten digits encoded in binary transmittance values ($\tau_0$ and  $\tau_1$). The lower transmittance, $\tau_0$, is varied, while $\tau_1=1$. Experimental points are reported for the quantum (blue) and classical (red) photon counting (PC) readout along with the theoretical predictions. The green band reports a simulation of the performance obtained with a pixel readout at the absolute classical limit. Panel \textbf{B} shows the classification error for almost ideal efficiency, $\eta=0.97$. Panel \textbf{C} reports the experimental quantum advantage, in dB ($\Delta P^{\mathtt{PC}} \defeq -10\log_{10}(P^{\mathtt{PC}}_{\mathtt{qua}}/P^{\mathtt{PC}}_{\mathtt{cla}})$).}\label{fig:kNN_ClasErr}
\end{figure*}

For the classification of the noisy digits we used a $k$-Nearest Neighbor ($k$-NN) classifier \cite{Cover_1967}. A $k$-NN classifier computes the distance between the pattern tested and all the patterns in its training set.  In the case of binary patterns, the most natural metric is the Hamming distance. The $k$ closest patterns of the training set to the tested pattern are selected, and the class of the test pattern is assigned as the more common among the $k$ closest. The experimental classification error is reported in Fig. \ref{fig:kNN_ClasErr}\textbf{A} as a function of the transmittance $\tau_0$, while in Fig. 
 \ref{fig:kNN_ClasErr}\textbf{B} we report the classification performance in the almost ideal ($\eta=0.97$) detection scenario. Note that the performance of the $k$-NN classifier is highly non-linear with respect to the single pixel error probability. For the values of $\tau_0$ corresponding to a classical error probability $p^{\mathtt{PC}}_{\mathtt{cla}}<0.3$ in Fig.~\ref{PxErr_X2}\textbf{A}, the classification is very robust to noise. In this range, the classification error is almost negligible for all the strategies and the quantum advantage is also rather small. However, for higher transmittance $\tau_0$, the algorithm becomes very sensitive to the readout noise level and here the advantage of quantum strategy is amplified. In fact, the experimental quantum advantage  in the pattern recognition task, represented in  Fig. \ref{fig:kNN_ClasErr}\textbf{C}, reaches up to 4 dB, which is much larger than the corresponding pixel error improvement.

At this point, the question arises: is the amplification of the quantum advantage we have observed for the $k$-NN classifier a property of this specific classifier, or a more general behavior characterizing the pattern classification task? To partially address this question, we repeat our analysis with another classification strategy that uses a refined machine learning approach. 

Convolutional neural networks (CNNs) are a category of feed-forward neural networks that are omnipresent in the field of image processing, and are particularly powerful tool for pattern recognition \cite{OShea_CNN,Gu_CNN}. They contain hidden layers that convolve input data from one layer to the next, in which the convolutions are performed according to parameterised filters (or kernels). The goal of learning is to optimise these filters in such a way that complex features can be reliably extracted and classified. This form of network architecture helps to minimise the number of network parameters and also helps the CNN to interpret features in a translationally invariant way, i.e.~qualities inferred in one area of an image can be reused elsewhere in the image. 

In this work, we apply CNN classifiers to the task of noisy pattern recognition on the MNIST dataset as before with the $k$-NN classifier, considering experimental input data from quantum-enhanced and classical sensors. We trained a collection of classifiers on a noisy training set in which each pixel in an image undergoes a bit flip with single-pixel error rate of $p_{\mathtt{train}}$, i.e.~the single-pixel training error. Meanwhile, the network parameter optimisation was carried out by backpropagation using a cross-entropy loss cost function. More precisely, we prepared a set of 16 CNNs, whose $p_{\mathtt{train}}$ was distributed across the interval $[0,0.5]$. The single-pixel training error is a particularly influential parameter as fine-tuning $p_{\mathtt{train}}$ helps classifiers become more robust to noise and learn how to reduce its impact on classification \cite{Nazar_NoisyCNN}. The classifiers could then be evaluated on the experimental test sets (constructed from $A_0$ and $A_1$) in order to quantify the performance of quantum-enhanced versus classical pattern recognition.

\begin{figure*}
	\vspace{0.10cm}
	\includegraphics[width=1\linewidth, angle=0]{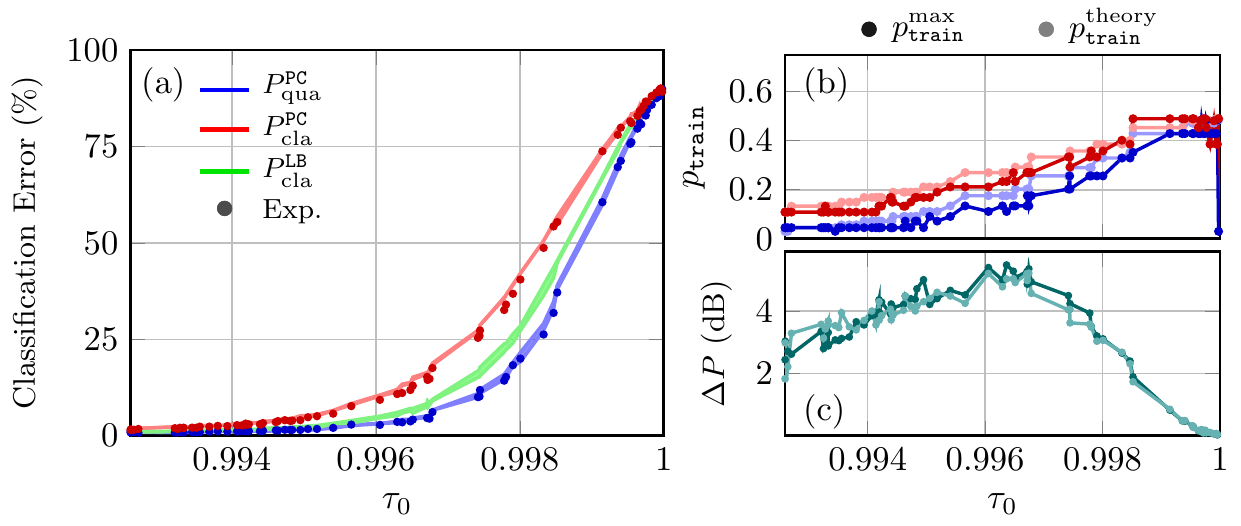}
    \caption{\textit{Quantum-enhanced pattern recognition via the CNNs.} Panel (a) displays the classification error of CNNs on the MNIST  using sensors with quantum resources (blue) or classical resources (red). We trained 16 CNNs on noisy training sets of 60000 images with single-pixel errors in the interval $p_{\mathtt{train}} \in [0,0.5]$. We used these CNNs to classify a test set of 10000 images simulated at different single-pixel transmissivities, $\tau_{0}$. The solid lines in (a) portray the simulated classification error region associated with a CNN using a single-pixel training error predicted by the corresponding theoretical single-pixel error bounds $p_{\mathtt{train}}^{\text{theory}}$ at $\tau_{0}$. Experimental data points are plotted for both quantum and classical photon counting protocols. Both the theory predicted value and the best-choice value for $p_{\mathtt{train}}$ are plotted in Panel (b). Finally, Panel (c) plots the experimental quantum advantage $\Delta P$ in decibels obtained with using TMSV states over classical coherent states (and photon counting). The dark points plot the advantage between CNNs that use the best-choice $p_{\mathtt{train}}^{\max}$ while the lighter points plot the advantage gained using theory predicted $p_{\mathtt{train}}^{\text{theory}}$.}
 \label{fig:CNN-ClassErr}
\end{figure*}

The results of the CNN strategy are presented in Fig.~\ref{fig:CNN-ClassErr}. Fig.~\ref{fig:CNN-ClassErr}\textbf{A} compares the performance of the classical and quantum-enhanced classifiers. We have a collection of CNN classifiers which have been trained using different single-pixel training errors, denoted by $\{ \textsf{C}[{p_{\mathtt{train}}}]\}_{{p_{\mathtt{train}}} \in [0,0.5]}$. It follows that some classifiers can garner better performance than others. Ideally, we would claim the classification error associated with the best classifier from our set, i.e.~$\textsf{C}[{p_{\mathtt{train}}^{\max}}]$. But this approach remains unrealistic, since it is not possible to predict \textit{a priori} what the best $p_{\mathtt{train}}$ would be. Instead, we may use the single-pixel error rates associated with the chosen discrimination protocol to motivate our choice of training error rate at a given transmissivity $\tau_{0}$, constructing the classifiers $\textsf{C}[p_{\mathtt{train}}^{\text{theory}}]$ (for more details, see Sec.~\ref{sec:NNC}). Fig.~\ref{fig:CNN-ClassErr}(b) plots these corresponding values against the single-pixel transmissivity.

Finally, Fig.~\ref{fig:CNN-ClassErr}(c) reports the quantum-advantage in the pattern recognition task (in decibels) showing a similar behaviour to the one observed in Fig.~\ref{fig:kNN_ClasErr}.\textbf{C}. This confirms that the quantum-advantage gathered through the single-pixel readout can be sustained throughout complex post-processing techniques, such as CNN training and evaluation.

\subsection{Parallel Multi-Pixel Readout} \label{sec:multipix}

While the scanning-like readout of the previous section represents a meaningful proof-of-principle necessary for a clean and faithful comparison with the theory, here we consider a more realistic scenario of a multi-pixel parallel readout, as it is depicted in Fig.~\ref{Scheme}\textbf{C}, where the pattern is acquired in single-shot.

The samples, containing digits from 0 to 9, were realized on a AR-coated glass slide combining laser lithography, sputtering deposition of titanium in high vacuum and lift off technique. The transmittance of the deposition is estimated as $\tau_0=0.987 \pm 0.003$. 
The parallel readout is doable since the state $\rho_\texttt{qua}$ produced by SPDC is spatially multimode. Pairwise correlated modes are detected by symmetric pairs of pixels in the CCD chip as represented in Fig.~\ref{Scheme}\textbf{C}, so that thousand pixel pairs are in principle non-classically correlated at the same time. However, to efficiently collect the correlated modes, the size of the pixel $l_{D}$ should be larger than the transverse cross-correlation length $l_{c}$ of the source. The consequence is a trade-off between the obtainable spatial resolution and the quantum advantage in the single pixel readout. We set the pixel size of our camera to fulfill the condition $l_{D}>l_{c}$ and at the same time to have a $28\times28$ image of the digits (more details are presented in Sec \ref{Materials&Methods}D).

\begin{figure*}
	\vspace{0.10cm}
	\includegraphics[width=1\linewidth]{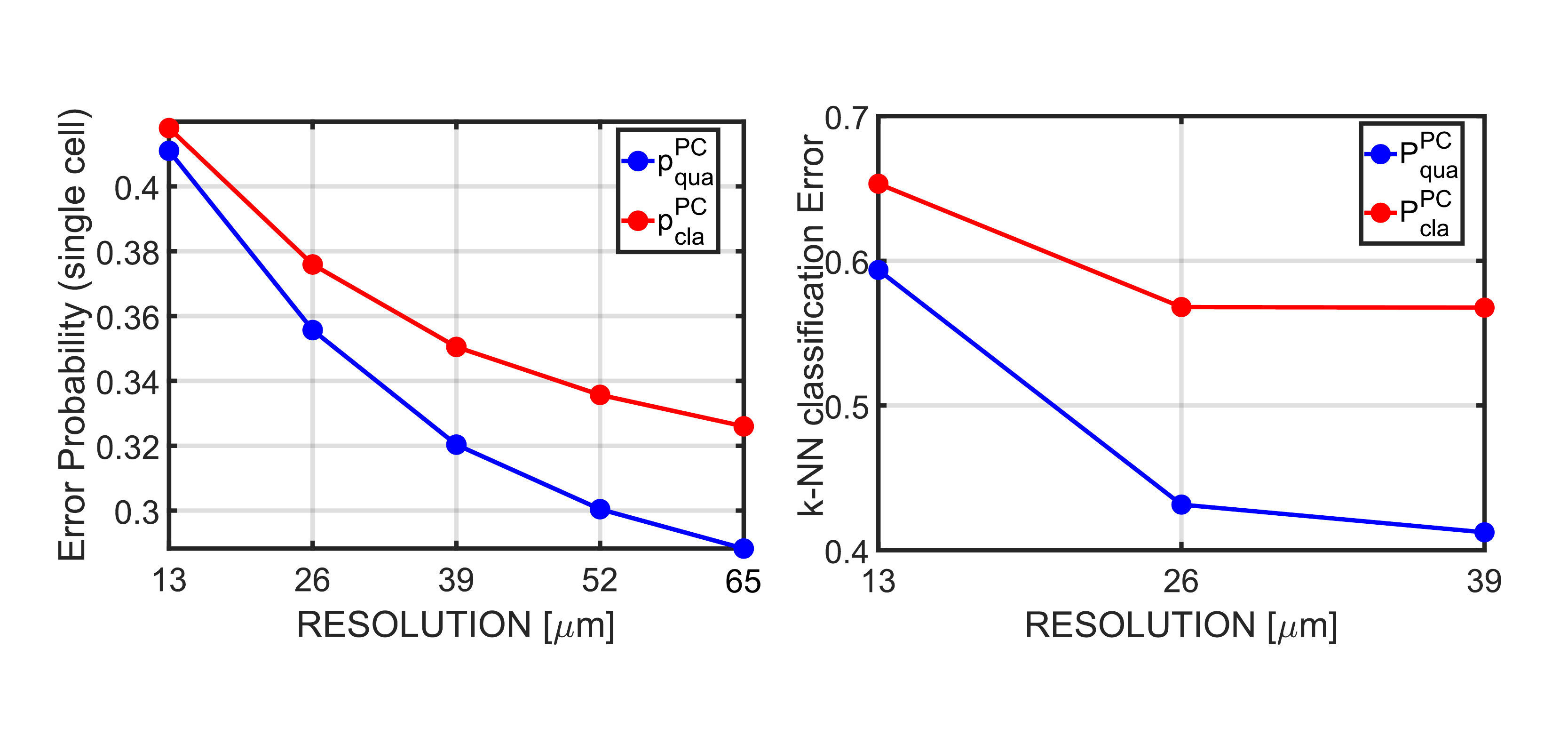}
	\caption{\emph{Multi-Pixel readout performance as a function of the spatial resolution.} Panel \textbf{A} shows the single-pixel error probability decreasing with the resolution at the object plane. The mean number of photons per pixel per frame, at full resolution (13 $\mu$m) is $\mu=1380 \pm 20$, while the total channel efficiencies of signal and idler branches are $\eta_S=0.795 \pm 0.01$ and $\eta_I=0.815 \pm 0.01$. The pattern transmittance level is fixed to $\tau_0=0.987 \pm 0.003$. Panel \textbf{B} reports the error in the pattern classification with a $k$-NN algorithm. \label{Multi-Pixel}}
\end{figure*}

Fig.~\ref{Multi-Pixel}\textbf{A} shows the experimental pixel error probabilities for both classical an quantum PC readout in function of the effective transverse resolution at the sample plane. The effective resolution is changed by applying a moving averaging filter to the acquired images, as described in Sec \ref{Materials&Methods}D, which has practically the same effect of increasing the pixel size. On one side, the increasing of the pixel size has the direct consequence of raising the number of signal photons per pixel, that explains the dropping of classical error probability, according to Eq.~(\ref{eq:4}). On the other side, as clearly shown in Fig.~\ref{Multi-Pixel}\textbf{A}, the quantum error probability decreases faster than the classical analogous, because the detectable quantum correlation increases while losing spatial resolution. 
Here, the comparison of the experimental results with the theoretical bounds is not straightforward and we will avoid it. In fact, when binary patterns are readout with finite resolution, for a detection pixel that falls across the edge between $\tau_0$ and $\tau_1$ the binary model is no longer valid. Such border effects are not negligible as the resolution decreases and become comparable with the the spatial features of the pattern. These features are very difficult to account for in a theoretical model. 

We evaluated the pattern recognition performance once again on the MNIST handwritten digit dataset using a $k$-NN algorithm for the classification.
While the test dataset is composed of 10000 patterns, producing and acquiring $10000$ samples is a not feasible and unnecessary task. In fact, the error in the classification can be broken down in two contributions. The first contribution is intrinsic to the classification algorithm and the second comes from the noise in the readout. In our case, with a pixel error larger than about 0.3, the classification error is dominated by the noise in sensing. For reference the classification error without noise is of the order of $4\%$, while in the noisy region it is clearly much larger. So, we assume that the classification error can be well studied by printing a small subset of patterns from the original MNIST, namely 10, and performing on each of them one thousand different acquisitions with independent noise realizations, forming a total dataset of 10000 images.

The resulting classification error is reported in Fig. \ref{Multi-Pixel}\textbf{B} as a function of the spatial resolution. The results are reported up to a resolution of 39 $\mu m$ which is equivalent to an effective reduction of the resolution of a factor 3 with respect to the original dataset.
Confirming the main results of the of the previous section, in a sensing scenario dominated by quantum noise, a relatively small quantum enhancement in the single pixel probability (namely from 0.35 to 0.32) can be amplified to around $20\%$ in the classification. 

\section{Discussion and Conclusion}

In this work we have experimentally demonstrated the advantage of quantum-enhanced sensing in the task of pattern recognition. As a test bed we have considered the problem of classification of handwritten digits. A quantum readout strategy based on TMSV states and photon counting measurement brings a relevant advantage in the classification errors with respect to any classical sensing. 

We have realized the experiment in two different configurations. In the first one, called ``scanning-like readout'', we have privileged the magnitude of the quantum advantage with respect to the acquisition time, since scanning of the sample is required. In this case we were able to clearly beat the optimal classical bound. In the second configuration, instead we have considered single-shot multi-pixel readout of the pattern. In this case we have shown a large advantage with respect to the classical bound that uses photon counting. 

Moreover, the presence of quantum advantage turns out to be independent on complexity of the classification algorithm. 
Having deployed both $k$-NN classifiers (a simple and robust strategy) and CNN classifiers (a machine learning enhanced strategy) on the same experimental data, similar results are obtained:
quantum advantage in the sensing is maintained and often amplified in the pattern recognition task.

The results presented here are particularly promising for short terms application in biological spatial pattern recognition. In general, they also pave the way for experimental studies of quantum enhanced pattern recognition in the spectral and temporal domains.

\section{Materials \& Methods}\label{Materials&Methods}

\subsection{Single pixel readout} \label{sec:SinglePix}
The single-pixel problem has been extensively analysed in Refs.\cite{Pirandola_2011, Invernizzi_2011, Nair_2011, Ortolano_2021, Ortolano_2022}. In the following we will compare three different local readout strategies. In the first scenario we consider classical input states $\rho_{\mathtt{cla}}$, i.e. a mixture of coherent states, paired with a theoretical optimal POVM $\Lambda$. The optimal probability of error in discriminating the outcome states $\rho_{\mathtt{cla}}(\tau_0)$ and $\rho_{\mathtt{cla}}(\tau_1)$ can be derived by the Helstrom bound~\cite{Helstrom_1976}, $(1-D(\rho_0,\rho_1))/2$, where $\rho_0$ and $\rho_1$ are two generic states, and $D(\rho_0,\rho_1)$ is the trace distance. Using the convexity of $D(\rho_0,\rho_1)$, Ref.\cite{Pirandola_2011} demonstrated the following bound to the classical error probability: 
\begin{equation}
	p^{\mathtt{LB}}_{\mathtt{cla}}=\frac{1-\sqrt{1-e^{-\mu_0(\sqrt{\tau_0}-\sqrt{\tau_1})^2}}}{2} \label{eq:2}
\end{equation}
where $\mu_0$ is the mean number of signal photons. As discussed in \cite{Pirandola_2011}, ancillary modes cannot enhance the performance of the readout with classical states, in fact the bound in Eq.(\ref{eq:2}) can be obtained by using a single mode coherent state transmitter $\rho_{\mathtt{cla}}=|\alpha\rangle \langle \alpha|$, with $|\alpha|^2=\mu_0$, as shown in Sec. \ref{OptMeas}.

However, the optimal measurement, that we discuss in Sec. \ref{OptMeas},  is purely theoretical, and no feasible receiver has been proposed yet to reach the classical lower bound in Eq.~(\ref{eq:2}). Thus, we consider a second, more practical scenario where the receiver is based on photon counting (PC) measurements that can be proven to perform close to the optimal one for practical purposes \cite{Ortolano_2021}.
The conditional photon number distribution at the receiver is $p(\textbf{n}|\tau_i)=\langle \textbf{n}| \rho(\tau_{i}) |\textbf{n} \rangle$, where $\textbf{n}=(n_S,n_I)$ is the number of measured photons in the signal and idler (the ancilla) systems. The optimal choice to recover the value of the bit $j$ is to choose it according to the Bayes' rule 
\begin{equation}\label{Bayes}
j=\argmax_{i}{p(\tau_i|\textbf{n})}.
\end{equation}
This condition is equivalent to the maximum likelihood decision $j= \argmax_i{p(\textbf{n}|\tau_i)}$ in the balanced prior probability case. The error probability with photon counting receiver is given by:
\begin{equation}
	p^{\mathtt{PC}}=\frac{1}{2} \sum_{\textbf{n}} \min_i p(\textbf{n}|\tau_i)\vspace{-3pt} \label{perrcla}
\end{equation} 
i.e., proportional to the overlap of the measurement outcome distributions $p(\textbf{n}|\tau_0)$ and $p(\textbf{n}|\tau_1)$. As for the optimal classical case, the conditional distribution overlap cannot be reduced by using ancillary mode, and the overlap is minimized for input signal states having the photon statistics of a coherent state, i.e. a Poisson distribution $\mathcal{P}_{\mu_0}(n)$, where $\mu_0$ is the mean number of photons~\cite{Ortolano_2021}. The statistics of the photons returning from a cell of transmittance $\tau_i$ still follows the Poisson distribution, in particular $\mathcal{P}_{\mu_0 \tau_{i}}(n_s)$. Substituting it in the conditional probabilities of Eq.~(\ref{perrcla}), the best performance achievable with classical states and PC measurements is \cite{Ortolano_2021}:
\begin{equation}
	p^{\mathtt{PC}}_{\mathtt{cla}}= \frac{1}{2}-\frac{\Gamma(\lfloor n^{th}+1\rfloor,\mu_0\tau_{0})-\Gamma(\lfloor n^{th}+1\rfloor,\mu_0\tau_{1})}{2 \lfloor n^{th}\rfloor!} 
	\label{eq:4} 
\end{equation}
where $\Gamma(x,y)$ is the incomplete gamma function, $\lfloor x \rfloor$ is the floor of $x$ and $n^{th}:=\mu_0 (\tau_1-\tau_0)/\log(\tau_1/\tau_0)$ is the intersecting point of the two Poisson distributions $\mathcal{P}_{\mu_0 \tau_{0}}$ and $\mathcal{P}_{\mu_0 \tau_{1}}$. 

The third strategy we consider here uses a collection of $M$ replicas of Two Mode Squezed Vacuum (TMSV) states in input, $\rho_{\mathtt{qua}}=| \text{TMSV} \rangle_{S,I} \langle \text{TMSV} |^{\otimes M}$, with $\vert \text{TMSV}\rangle_{I,S}\propto\sum_{n} \sqrt{\mathrm{P}_{\mu_0}(n)}\vert n\rangle_{S}\vert n\rangle_{I}$ and $\mathrm{P}_{\mu_0}(n)=\mu_0^n/(\mu_0 + 1)^{n+1}$ being a thermal distribution with $\mu_0$ mean photons. The joint probability of signal and idler photons of a TMSV state after the interaction with the cell of transmittance $\tau_i$ is $p_{\mu_0}(n_S,n_I|\tau_u)= \mathrm{P}_{\mu_0}(n_I) \mathcal{B}(n_S|n_I,\tau_i)$, where $\mathcal{B}$ stands for the binomial distribution. For the multi mode state $\rho_{\mathtt{qua}}$ the joint distribution is simply obtained by substituting the single mode thermal statistics $\mathrm{P}_{\mu_0}(n_I)$ with a multi-thermal one, $\mathrm{P}_{\mu_0, M}(n_I)$, obtained by the convolution of $M$ thermal modes. The conditional joint probability evaluated in this way can be substituted in Eq.~(\ref{perrcla}) to obtain the error of the quantum strategy, $p^{\mathtt{PC}}_{\mathtt{qua}}$. This quantity can be evaluated numerically. It turns out that, for a fixed total number of photon transmitted to the cell $\mu_0 M$, the most effective way to exploit quantum correlation is to span the photons in a large number of modes so that $\mu_0 \ll 1$ and $M \gg 1$. Moreover, the local multi-thermal distribution approaches the Poisson one. This is the regime that we consider in the simulation, and the one that is realized in the experiment.

\subsection{Optimal local receiver for classical states}
\label{OptMeas}
The single pixel readout probability of error that can be reached with classical states, having a fixed number of signal photons $\mu$, is lower bounded by the limit in Eq.~(\ref{eq:2}). As previously mentioned, this limit can be saturated by using a single-mode coherent state without idlers. This can be seen using the fact that a pure loss channel, with transmissivity $\tau$, maps a coherent state $|\alpha\rangle$ into another, amplitude damped, coherent state $|\tau \alpha \rangle$. Once the input state $\rho_{in}$ is fixed, the discrimination of two channels having different transmissivity is reduced to a binary quantum state discrimination between the two possible outputs, $\rho_0$ and $\rho_1$ for which the lowest error probability is given by the Helstrom formula:
\begin{equation}
p^{\mathtt{H}}= \frac{1}{2} (1-||\pi_0 \rho_0 - \pi_1 \rho_1||), \label{eq:HB}
\end{equation}
where $||\cdot ||$ is the trace distance and $\pi_0,\pi_1$ are the prior probabilities.

For coherent input, the potential outputs of the channel are pure and their trace distance can be expressed in terms of the overlap $\zeta=\langle\sqrt{\tau_0}\alpha|\sqrt{\tau_1}\alpha\rangle=e^{-\frac{\mu}{2}(\sqrt{\tau_0}-\sqrt{\tau_1})^2}$ as:
\begin{equation}
|| \pi_0|\sqrt{\tau_0}\alpha\rangle\langle\sqrt{\tau_0}\alpha| - \pi_1|\sqrt{\tau_1}\alpha\rangle\langle\sqrt{\tau_1}\alpha||| = \sqrt{1-4 \pi_0 \pi_1 |\zeta|^2}. \nonumber
\end{equation}
Substituting this in Eq.(\ref{eq:HB}) and setting equal priors $\pi_0 = \pi_1 = 1/2$ we get:
\begin{equation}
p_{\mathtt{coh}}=\frac{1-\sqrt{1- |\zeta|^2}}{2},
\end{equation}
which coincides with $p_{\mathtt{cla}}^{\mathtt{LB}}$ proving that a single mode coherent state saturates the classical bound (originally proven for an arbitrary P-representation~\cite{Pirandola_2011}).

The optimal probability of error is achieved by projecting into the eigenstates of the Hermitian operator:
\begin{equation}
\Lambda= \frac{1}{2}(|\sqrt{\tau_0}\alpha\rangle\langle\sqrt{\tau_0}\alpha| - |\sqrt{\tau_1}\alpha\rangle\langle\sqrt{\tau_1}\alpha|).
\end{equation}
We can find an orthogonal basis to represent $\Lambda$ by performing the Gram-Smith orthogonalization  on the pair of vectors $(|\sqrt{\tau_0}\alpha\rangle,|\sqrt{\tau_1}\alpha\rangle)$ spanning its support. So we define the basis:
\begin{align}
	|0\rangle= |\sqrt{\tau_0}\alpha\rangle && |1\rangle= \frac{|\sqrt{\tau_1}\alpha\rangle -\zeta|\sqrt{\tau_0}\alpha\rangle}{\sqrt{1-|\zeta|^2}} \label{eq:bs}
\end{align}
The eigenvectors of $\Lambda$ can be computed with simple algebra to be, in terms of this basis:
\begin{align}
|+\rangle&= \frac{1}{\sqrt{2}}\sqrt{1-\sqrt{1-\zeta^2}}|0\rangle - \frac{1}{\sqrt{2}}\sqrt{1+\sqrt{1-\zeta^2}}|1\rangle \\ 
|-\rangle&= \frac{1}{\sqrt{2}}\sqrt{1+\sqrt{1-\zeta^2}}|0\rangle + \frac{1}{\sqrt{2}}\sqrt{1-\sqrt{1-\zeta^2}}|1\rangle 
\end{align}
And substituting the definitions in Eq.(\ref{eq:bs}) leads to:
\begin{align}
	|+\rangle&= \sqrt{\frac{1-\sqrt{1-\zeta^2}}{2(1-\zeta^2)}}|\sqrt{\tau_1}\alpha \rangle -\sqrt{\frac{1+\sqrt{1-\zeta^2}}{2(1-\zeta^2)}}|\sqrt{\tau_0}\alpha\rangle \label{eq:03} \\
	|-\rangle&= \sqrt{\frac{1+\sqrt{1-\zeta^2}}{2(1-\zeta^2)}}|\sqrt{\tau_1}\alpha\rangle - \sqrt{\frac{1-\sqrt{1-\zeta^2}}{2(1-\zeta^2)}}|\sqrt{\tau_0}\alpha \rangle 
\end{align}
 The bit can then be recovered by measuring over the projectors $\Pi_0=|+\rangle \langle +|$ and $\Pi_1=|-\rangle \langle -|$. Once the measurement is performed the value $\tau_0$ is selected if the outcome is $+$ and $\tau_1$ is selected otherwise. This measurement paired with a coherent state input would in theory saturate the classical bound in Eq.(\ref{eq:2}). However, as pointed out in the main text, from a practical point of view an implementation of this kind of measurement would be very complicated, if feasible at all, with current technology.
 
\subsection{$k$-NN classification \label{sec:KNN}}
Let $\bs{i}_0=\{b^{(0)}_{ij}\}$ and $\bs{i}_1=\{b^{(1)}_{ij}\}$, $b_{ij}=0,1$, be two binary  images of dimension $d\times d$. The Hamming distance, $d_\mathtt{H}$, between $\bs{i}_0$ and $\bs{i}_1$ is defined as:
\begin{equation}
d_\mathtt{H} (\bs{i}_0,\bs{i}_1)=\sum_{i=1}^d \sum_{j=1}^d |b^{(0)}_{ij}-b^{(1)}_{ij}|.
\end{equation}
The $k$-NN is a supervised classification method that uses the known label of images in a training set $\mathcal{T}$ to assign a class to images in a test set $\mathcal{V}$. Given an image $\bs{i}_v$ in $\mathcal{V}$, the set of its $k$ closest images in terms of the hamming distance, $\mathcal{K}=\{\bs{i}_\mathcal{T}^{(1)},..., \mathbf{i}_\mathcal{T}^{(k)}\} \subset \mathcal{T}$, are selected from $\mathcal{T}$.  The image $\bs{i}_v$ is assigned to the class $c_v$ which is the most common among the images in $\mathcal{K}$. The most intuitive case is given by the $1$-NN classifier, that assigns to the image $\bs{i}_v$ the same class of its closest image in the training set $\mathcal{T}$.

\subsection{Neural Network classification\label{sec:NNC}}

Consider a dataset of images $\mc{I} \defeq \{ \bs{i}_j; c_j \}_j$ in which each image $\bs{i}_j$ possesses a classification (or label) $c_{j}$. For the task of classification, a neural network \cite{Nielsen_NN, Murphy_ML} is a universal approximating function $f_{\bs{\theta}}$ (where $\bs{\theta}$ denotes the parameter set of the neural network) which takes an image as input and provides an approximation of its classification as an output, $f_{\bs{\theta}}(\bs{i}_j) = c_{j}^{\prime}$. Training a neural network as a classifier equates to optimising its parameters $\bs{\theta}$ such that $f_{\bs{\theta}}(\bs{i}_j) \approx c_{j}$ becomes a reliable classifier over the entire dataset. That is, the goal of training is to maximising the accuracy of the classifier over the training set in such a way that it can generalise to new, unforeseen samples. In a neural network, the parameter set $\bs{\theta}$ describes correlations between layers of neurons, capable to encoding highly complex relationships between input and output. 

When building neural network classifiers for which there already exists experimental data (as in the case of MNIST) supervised learning is a very effective strategy. This involves utilising a set of images whose classifications are already exactly known, and training the neural network on this data. In this way, the classifier can be optimised reliably and extract meaningful features from the dataset in order to learn correlations between input patterns and their classification. The classification error of the classifier is measured against an evaluation (or test) set; a set of images which the network has never seen during training and therefore must extrapolate. Given a $K$-element evaluation set of images and labels $\mc{V} = \{ \bs{i}_k; c_k \}_{k=1}^{K}$ and a classifier $f_{\bs{\theta}}$ then the classification error is approximated as
\begin{equation}
P_{f_{\bs{\theta}}} = \frac{1}{K} \sum_{k=1}^K \delta[f_{\bs{\theta}}(\bs{i}_k), c_k]
\end{equation}
where $\delta[f_{\bs{\theta}}(\bs{i}_k), c_k]$ is a Kronecker-delta function that returns unity iff the classifications are the same, otherwise it will return zero. 

In this work, we exploit CNNs as classifiers upon on our experimental dataset. Using the Julia Flux package \cite{Flux_2018a,Flux_2018b}, we construct CNNs with three convolutional layers and a dense layer, optimising a cross entropy loss function during training. This produces classifiers of sufficient accuracy to utilise in this study. We use the MNIST handwritten digit dataset, composed of a training set $\mc{T}$ with $6\times 10^4$ images and an experimental evaluation set $\mc{V}$. The experimental dataset is clearly unchangeable and is inherently noisy. But the initially noiseless training set can be manipulated to modify and enhance our classifiers. As discussed in the main text, it is useful to train a CNN on noisy data if it is expected to classify noisy data. To do this, we simulate noisy training sets $\mc{T}_{p_{\mathtt{train}}}$ in which each pixel in every image is independently bit flipped according to the probability $p_{\mathtt{train}}$. We then trained a collection of different classifiers using different values of $p_{\mathtt{train}} \in [0,0.5]$ and evaluated them on the experimental data. 

It is difficult to know what the best case $p_{\mathtt{train}}$ is with respect to the transmissivity of the sampled pixels. As depicted in Fig.~\ref{fig:CNN-ClassErr} we can identify a best choice classifier from our collection as that which minimises its classification error with respect to the evaluation set. However, it is not practical to select a best choice classifier \textit{a posteriori} (this would defeat the purpose of discrimination). Instead, we present the best-case performance for benchmarking purposes and more practically motivate our chosen $p_{\mathtt{train}}$ by mapping the single pixel transmissivity to a single-pixel error probability associated with the discrimination protocol being used.

\subsection{Experimental Setup and Noise Reduction} \label{sec:ExpSet}
\begin{figure}
	\vspace{0.10cm}
	\includegraphics[width=\columnwidth]{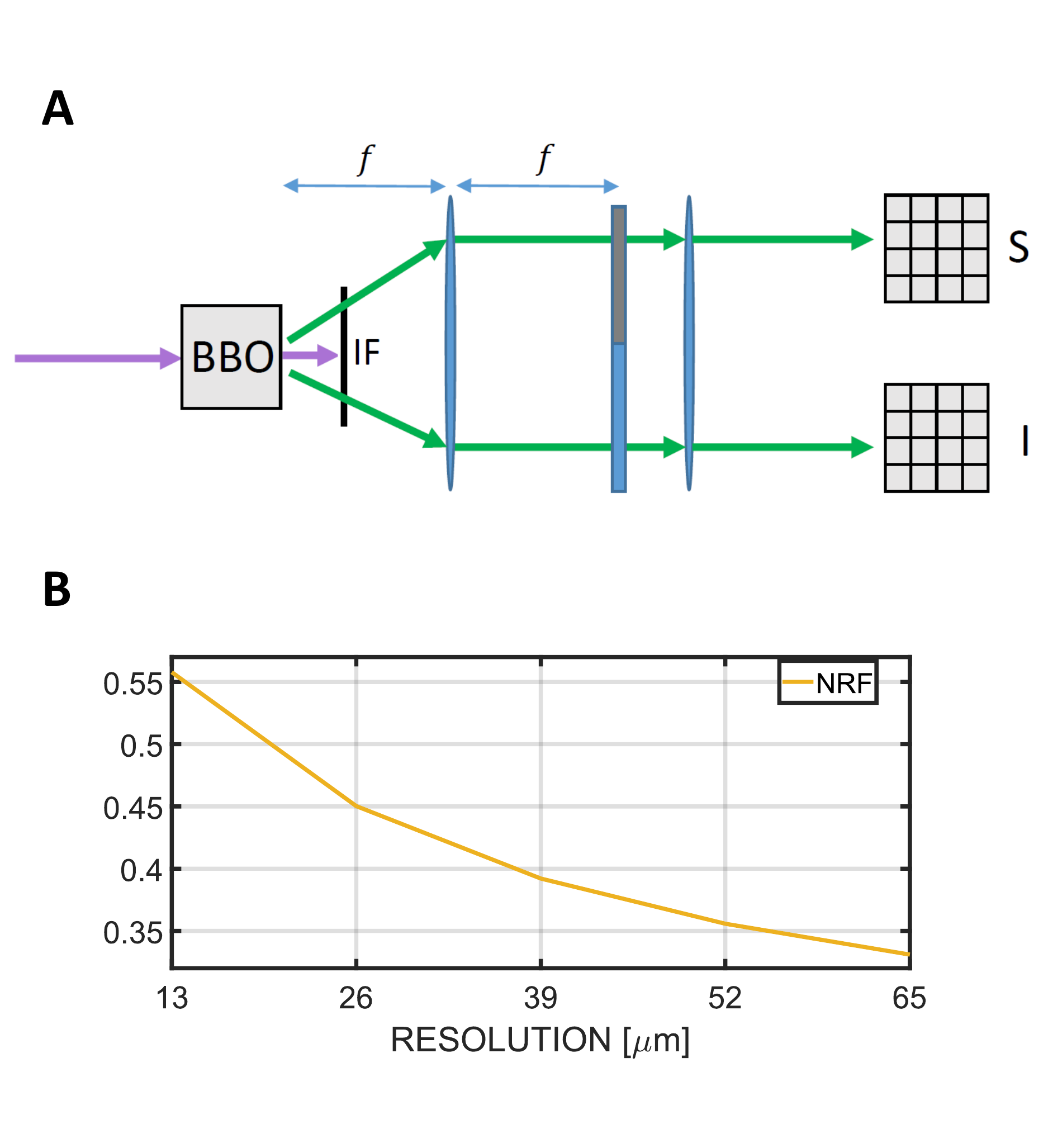}
	\caption{\label{expscheme} \textbf{A} Schematic of the experimental setup described in Sec. \ref{sec:ExpSet}. \textbf{B}. Noise reduction factor (NRF) as a function of the effective spatial resolution.}
\end{figure}

Acquisitions for both the experimental configurations reported in Fig. \ref{Scheme}\textbf{B-C}, are performed using the setup sketched in Fig. \ref{expscheme}. It exploits SPDC to generate multi-mode pairwise correlated collections of TMSV states, by pumping a type II Beta Barium Borate (BBO) non-liner crystal with a 100 mW continuous laser at wavelength $\lambda_p=405$ nm. The SPDC photons are selected around the degenerate wavelength $\lambda=2\lambda_p=810$ nm by an interference filter (IF) at $(800 \pm 20)$ nm. The correlations in momentum at the source are converted in spatial correlations at the "object" plane, by use of a lens having focal length $f=1 \text{cm}$, in an $f-f$ configuration, performing a Fourier transform on the incoming field. The test patterns to be classified are physically realized as depositions, according to the procedure of Sec. \ref{hdd}. The patterns are placed in the signal beam at the object plane, while a blank glass slide is placed on the idler beam to match the two optical paths. The object plane is then imaged to the charged coupled device (CCD) camera using a second lens, with magnification of about 8 times. The camera (Princeton Instrument Pixis 400BR Excelon) has a nominal quantum efficiency $>95\%$, electronic noise of 4$e^{-}/\text{pix}/\text{frame}$ and the physical pixel is of 13 $\mu$m. The total channel efficiencies $\eta_S$ and $\eta_I$, are estimated close to 80\% according to the procedure outlined in Ref.s \cite{Brida_20100, Avella_2016} 

In the scanning-like configuration, we introduce in the signal beam a variable absorber that is fine tuned in the range $1-\tau_{0}=10^{-2}\div10^{-3}$. Signal and idler photons $n_{S}$ and $n_{I}$ are obtained integrating over two corresponding wide spatial areas composed of a large number of physical pixels, becoming equivalent to two bucket detectors. This leads to the most efficient collection of the correlated photons \cite{Ortolano_2021b} and thus to reach an higher quantum advantage, as we will explain in the following.

In contrast, in the parallel multi-pixel readout, we exploit the spatial resolution of the detector to have a one-shot (without scanning) image of size $28\times 28$ of a test digit in the signal arm, which is the original dimension of the binary images of the MNIST dataset. For that, we perform a $8\times8$ hardware binning of the physical pixels, to obtain an effective readout pixel of $l_{D}=104\mu$m, corresponding to a resolution of $l_{R}=13\mu$m at the object plane (not to be confused with the original physical pixel size of the detector  mentioned earlier). 

However, a trade-off exists between spatial resolution and quantum noise reduction achieved exploiting the signal-idler correlations. Actually, correlated photons from the multi-mode SPDC arrive at the detector within a certain transverse spatial uncertainty, that comes from the finite size of the pump beam. For this reason, the signal-idler intensity cross-correlation is well approximated by a Gaussian function with finite coherence length $l_c$  at the detector plane (in our case is $l_c\sim40\mu$m). To efficiently detect the correlated modes, the pixel size, or more in general the detector resolution area, must be larger than this coherence length, $l_D\geq l_c$ \cite{Meda_2017}. In Sec.(\ref{sec:multipix}) we analyse the patter recognition performance as a function of the resolution. For a fair comparison in the classification, it is more suitable to have final binary images with always the same number of pixels. Thus, instead of changing the resolution by performing a simple pixel binning, we preferred to change it by applying an averaging filter of appropriate size. The averaging filter substitutes to each pixel count, the average of the counts in a neighborhood of size  $d\times d$. This is repeated for each original pixel of the  $28\times 28$ matrix, so that the final size of the image in terms of pixels is unvaried, although the effective spatial resolution is reduced. In the graph of Fig. \ref{Multi-Pixel}\textbf{A}, the parameter $d$ ranges between 1 and 5, which correspond to an effective resolution $l_{R}$ between $13\mu$m and $65\mu$m.  The improvement of the quantum correlations in function of the image resolution can be witnessed by the noise reduction factor (NRF), defined as $\text{NRF}=\langle \Delta^2 (n_S - n_I) \rangle/\langle n_S + n_I \rangle$  \cite{Jedrkiewicz_2004,Bondani_2007, Blanchet_2008, Agafonov_2010, Perina_2012}. The $\text{NRF}$ is an indicator of non-classical correlations, since it can show values in the range $0<\text{NRF}<1$ only for non-classical fields, while it is $\text{NRF}\geq1$ for classical light. In Fig.(\ref{expscheme}.\textbf{B}) we report the measured NRF as a function of the effective resolution at the object plane.

\subsection{Handwritten digit deposition} \label{hdd}
The CAD of the 10 digits (from 0 to 9) was realised by normalising the dimension of each binary images (28x28 pixels) of digit in squares of maximum dimension of 400x400 $\mu \text{m}^{2}$ spaced of 1000 $\mu$m. A laser lithography process was performed with Heidelberg Instruments uPG101 system equipped with a UV laser source at 375 nm. The deposition of titanium was carried out by sputtering in high vacuum at low rate (0,08 nm/s). Thicknesses from 2 to 4 nm were deposited to obtain a transmittance value suitable for the experiment. Final patterning was obtained removing titanium excess through lift off technique.

\section*{Acknowledgments}
 Part of this work (sample fabrication) has been carried out at QR Laboratories, INRiM, a micro and nanofabrication lab. IRB thank the INRIM reasearcher Matteo Fretto for supervising the sample realization.
\subsection*{Funding}
This work was founded by the EU via ``Quantum readout techniques and technologies'' (QUARTET, Grant agreement No 862644). 
\subsection*{Author contributions}
GO and IRB devised the present realization of quantum enhanced patter recognition, with advice of SP. EL, PB, CN and GO performed the experimental acquisitions. GO did the data analysis on the experimental data and computed the optimal local measurement to saturate the classical lower bound. GO and CH performed the classification analysis for $k$-NN and CNN respectively. The samples have been prepared by GL. MG, head of the INRIM quantum optics sector and IRB, supervised the project. GO, IRB, SP, CH wrote the paper with the contribution of all authors.

\subsection*{Competing Interests}
The authors declare no competing interest.
\subsection*{Data availability} All data needed to evaluate the conclusions are reported in the paper. Further data, for reproducibility of the results, will be available in a public repository linked to the published paper.

\bibliography{bib}

\end{document}